\documentstyle[prl,aps,multicol,epsf]{revtex}

\begin{document}
\draft
\tighten

\title{ Quantum phase transitions in cuprates: stripes and antiferromagnetic 
supersolids.}
\author{J. Zaanen}
\address{Institute Lorentz for Theoretical Physics, 
Leiden University, P.O. Box 9506, NL-2300 RA Leiden, The Netherlands}
\date{\today}
\maketitle

\begin{abstract}
It is believed that the magnetic fluctuations in cuprate superconductors
reflect the proximity to a quantum phase transition. It will be
argued that this notion acquires further credibility if combined with 
the idea that the superconducting state is  in a tight competition 
with the stripe phase over a large range of hole concentrations.
On basis of existing data and some simple considerations,
a zero temperature phase diagram will be proposed with an unusual topology 
which is unique to the competition stripe phase-superconductivity. It is
argued that the existence of a state which is at the same time stripe
ordered and superconducting (antiferromagnetic supersolid)
is a prerequisite for quantum critical behavior in the magnetic sector.
Various predictions follow which can be tested experimentally.
\end{abstract}


\begin{multicols}{2}
\narrowtext

\section{Introduction}

Untill not long ago, it was assumed  that cuprate physics 
was about a rather anomalous metallic state, subjected to a superconducting
instability, and a Mott-insulating antiferromagnetic state
 in a remote corner of the phase diagram.  A  consequence
of the discovery of the stripe phase\cite{tran}
 is that stripes have to be added to
the list of states which compete at zero temperature. Although still
littered with uncertainties, enough experimental information is
available to conjecture the general shape of the 
zero-temperature ($k_B T = 0$) phase diagram: see Fig. 1. 
The $x$ axis has the usual meaning of hole concentration and the 
other axis  is taken in a rough sense as an influence which helps
charge localization over superconductivity : I call this $g^{-1}$ since
it is similar to the inverse of the coupling constant of a quantum
phase-dynamics problem. 

\begin{figure}[h]
\hspace{0.00 \hsize}
\epsfxsize=0.9\hsize
\epsffile{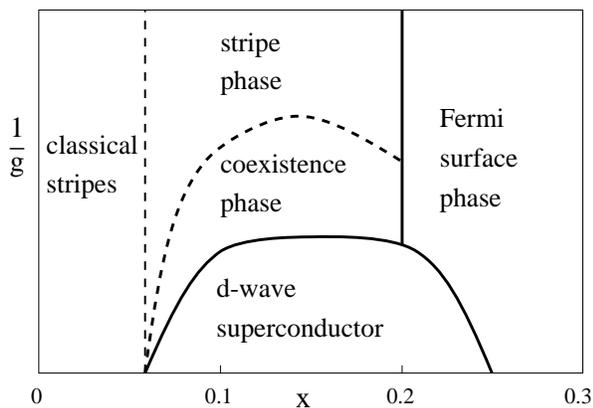}
\caption{ The topology of the zero-temperature phase diagram of high $T_c$
superconductors, as function of doping ($x$) and a control parameter ($g^{-1}$) 
surpressing superconductivity and/or promoting the stripe phase (magnetic
fields, the LTT deformation, $Zn$ doping).}  
\label{f1}
\end{figure}

As will be further discussed in section II, it is about a metal
competing with the superconductor at high dopings, about presumably some
nickelate-like `classical' stripes at very low dopings which are strongly
affected by quenched disorder and, last but not least,
by an `underdoped regime' 
where over a large concentration range the superconductor competes with
the stripe phase. Although still quite controversial, it might be that
at intermediate $g^{-1}$ stripes and superconductivity coexist
in this underdoped regime\cite{tran1}.  A main
aim of this contribution is to analyze the role of this `coexistence' or,
more precisely, `antiferromagnetic supersolid' phase. 

At stake is that zero temperature (`quantum') phase transitions can 
govern the physics  at finite times, lengths and temperatures,
also if one is away from the locus of the transition in $T=0$ parameter
space. If the transition is continuous, and if the competition can be
described in terms of a bosonic field theory, one meets the phenomena
often referred to by quantum criticality\cite{sachdev}. 
A generic property of the quantum critical regime 
is that the phase-relaxation time $\tau_{\phi} \simeq \hbar / k_B T$,
while for an effectively Lorentz invariant dynamics the correlation
length $\xi$ is related to a geometrical average of 
energy $\omega$ and temperature: $1 / \xi^2 \sim (\hbar \omega)^2 +
(k_B T)^2$\cite{sachdev}. 
Using inelastic neutron scattering, Aeppli {\em et al} 
demonstrated recently that this scaling behavior 
is obeyed by the incommensurate magnetic fluctuations of
$La_{1.85} Sr_{0.15} CuO_4$ in its normal state\cite{aeppli}.
Since these fluctuations
are found at the same wavenumbers as the magnetic superlattice Bragg peaks
of the static stripe phase\cite{tran1,yameps}, 
it is tempting to think that these fluctuations
have to do with the proximity of the stripe antiferromagnetic order. This
interpretation is further helped by the observation that the spectrum of
incommensurate fluctuations acquires a gap at low temperatures, and this
gap $\Delta_{\varepsilon}$ is very small ($6 meV$) as compared to the
lattice scale exchange ($100 meV$)\cite{yamada}: 
the smallness of this gap signals 
the close proximity to the quantum critical point. In addition, it has been
argued that the antiferromagnet found in the LTT cuprates
$La_{2-x-y} RE_y Sr_x CuO_4$ ($RE = Eu, Nd, ...$) is characterized by
strong quantum fluctuations\cite{vanduin}, indicating the proximity
of the stripe antiferromagnet itself to the quantum disordering transition. 

The above interpretation points at the presence of a second order quantum
phase transition, at least involving the spin sector. As I will show,
this observation together with the phase diagram of Fig. 1 puts some
strong constraints on the form of the effective low energy theory. 
The argument rests on: (a) Some well established notions developed 
in the context of the strongly interacting boson problem, centered
around the concept of supersolid order\cite{supers,otterlo,balents}. (b)
a straightforward extension to the $T=0$ (quantum) case of   
the phenomenological theory by Zachar, Kivelson and Emery\cite{zachar} 
for stripe order.  These matters will be discussed in section III.       
Since the phasediagram Fig. 1 has to my knowledge not been proposed 
before, let me first discuss its somewhat uncertain status.

\section{Topology of the zero-temperature phasediagram.}

The assumption underlying the construction of Fig. 1 is that 
the `perturbations' stabilizing stripe order can all be understood
as the `$g^{-1}$' (y-axis) of Fig. 1. This is not quite obvious,
and even if true, the physically realizable $g^{-1}$ parameters
are not well behaved, with the effect that big portions of the
phase diagram have not yet been accessed. The best documented `$g^{-1}$' is
the rare earth concentration $y$ in the cuprates of composition
$La_{2-x-y} RE_y Sr_x CuO_4$ ($RE = Nd, Eu, \cdots$) showing the low
temperature tetragonal (LTT) distortion. 
As argued by Tranquada {\em et. al.}\cite{tran}, 
the LTT deformation can be
regarded as a relatively weak collective pinning potential. If this
potential could be switched on continuously, it would be close to an
ideal realization of $g^{-1}$. The problem is, however, that at a
critical substitution $y_c$ the LTT deformation switches on 
in a first order transition\cite{buechner}
 as expected for a 3D structural transition.
Apparently, this corresponds with a jump from deep inside the superconducting 
regime into the coexistence regime of Fig. 1. 
A next candidate is substitution by impurities like $Zn$\cite{yamadazn}. 
A problem is that 
this introduces additional quenched disorder into the problem, further
obscuring the clean limit physics\cite{castronebal,castronetex}. Finally,
magnetic fields\cite{boebinger}
are believed to stabilize stripes as well. Besides the
practical problem that few experiments can be done in $\sim 60$ T fields,
additional complexities are expected here as well\cite{notevort}. It is 
a matter of high priority for the experimental community to search for
alternative $g^{-1}$ like control parameters.

Given these reservations, the phase diagram topology follows directly
from experiments. The metal-insulator transition at $x \simeq 0.20$,
as seen in magnetic fields\cite{boebinger}
(and $Zn$ substitution experiments\cite{uchida})
coincides with the concentration where Tranquada {\em et. al.}
find the stripe order parameter to disappear in the LTT system\cite{tran1}. 
It is
firmly established that in the concentration range $x=0.125 - 0.20$
incommensurate magnetic order is present in the LTT system\cite{tran1}
 and some evidence 
is available for the presence of this order at $x < 1/8$, even in
$La_{2-x} Sr_x CuO_4$\cite{budnik} itself. 
A second singular doping concentration is
$x \simeq 0.06$ where the superconductivity disappears. Remarkably, 
Yamada {\em et. al.}\cite{yameps}
 find that with the diminishing of the superconductivity
also the incommensurate magnetic fluctuations disappear, being replaced 
by a broad peak centered at the $( \pi/a, \pi/a )$ wavevector. Although
evidence exists showing that one or the other collective phenomenon
involving the holes and the spins is at work in the doping range $0 <
x < 0.06$\cite{hammel}, 
it remains to be seen if this is related to the stripes at
higher doping. Finally, a crucial issue is whether the superconductor
and the stripe phase are separated by an intervening microscopic
coexistence phase: see Fig. 1. The experimental situation\cite{tran1}
is far from
settled, and a main purpose of this communication is to discuss the
possible role of this coexistence phase. However, assuming that it
exists, it is clear that for increasing `$g^{-1}$' the superconductivity
will eventually vanish. It has been shown that for increasing LTT
tilt angle in the $La_{2-x-y} Sr_x Nd_y CuO_4$ system a region opens
up around $x = 1/8$ which is not superconducting\cite{buechner}. 

The novelty of the phase diagram, Fig. 1, is that as function of 
doping {\em lines} of $T=0$ phase transitions
are present, instead of the isolated points which are discussed in the
theoretical literature. It is experimental fact that stripe phases exists
in a large doping range\cite{tran1}. Different from Mott-Hubbard insulators, 
the charge- and spin order exists away from points of low order 
charge commensuration. Although the ordering seems characterized by a 
partial commensuration\cite{liqcr}, what matters in first instance
is that stripes can be formed in a large range of dopings. Because
superconducting order is not critically dependent on the hole density either, 
quantum phase transitions can occur over a wide range of dopings.  
This helps to remove a standard difficulty associated with the 
idea that high $T_c$ superconductivity is related to
the physics of quantum phase transitions. The quantum criticality as referred
to in the introduction is apparently present over a large doping range, and
this is not natural if the physics is controlled by an isolated quantum
critical point on the doping axis. However, it becomes more natural
given that there is a line of critical points as function of doping.

Ignoring the low doping regime, in addition to the line of stripe related 
transitions there is a single isolated singular point as function of doping: 
the metal-insulator transition at $x \simeq 0.20$. The $T=0$ phasediagram
of Fig. 1 actually suggests a particular interpretation of
the {\em finite} temperature cross-over diagram as constructed by Pines and
coworkers\cite{pines},  based on the analysis of a vast amount of data. This
cross-over diagram is reproduced in Fig. 2. The spin-gap temperature $T^*$ 
(dashed line) can be
interpreted as measuring the `distance' between the superconductor and
the coexistence phase. If temperature exceeds the spin gap the `$z=1$'
quantum critical regime is entered, which is associated with the freezing
of the stripe antiferromagnetism. It is obvious that this spin gap will,
at least initially, grow as function of increasing doping. However, there
are also crossover lines associated with the singular $x=x_{MI}$ of
the metal insulator transition: $T_{cr}$ (full lines)\cite{cambridge}. 
In sharp contrast with
$T^*$, $T_{cr}$ is strongly doping dependent, as expected for a cross-over
line associated with an isolated point on the doping axis.

\begin{figure}[h]
\hspace{0.00 \hsize}
\epsfxsize=0.9\hsize
\epsffile{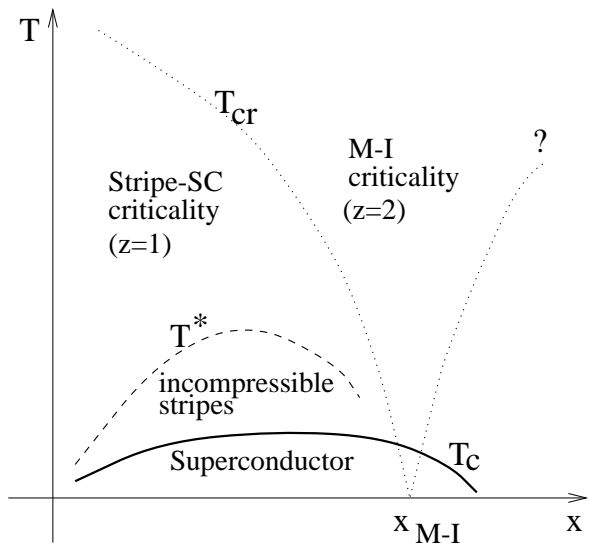}
\caption{ Finite temperature crossover diagram according to 
Pines {\em et. al.} [22],
but now with an interpretation motivated by  the zero
temperature phase diagram, Fig. 1. The dotted lines ($T_{cr}$)
indicate the cross-over
to the quantum critical regime controlled by the metal-insulator transition.
To the underdoped side, a regime is entered below $T_{cr}$
which is controlled by the
line of quantum phase transitions from the superconductor to the
coexistence phase. The spin-gap temperature $T^*$ measures the 
$T=0$ `distance' to the coexistence phase.}   
\label{f2}
\end{figure}

The precise nature of the critical regime associated with the metal-insulator
transition depends  on the nature of the metal in the overdoped
regime. Assuming that this metal is a Fermi-liquid, the critical regime
is likely of the Millis-Hertz variety\cite{millis}
as controlled by the vanishing of
the stripe order. Such an interpretation acquires further credibility by the 
observation that this regime is characterized by mean field exponents 
($z=2$), and the remaining issue is if the transition is dominated by
the spin-channel\cite{pines} or the charge channel\cite{castellani}. 
Obviously, it remains to be seen if high $T_c$ superconductivity has
anything to do with Fermi-liquid physics\cite{anderson}.

\section{Criticality and the antiferromagnetic supersolid.}

Let us now focus on the doping regime characterized by the competition
between superconductivity and the stripe phase. It is assumed that the
long wavelength dynamics is governed by conventional bosonic ordering 
fields, described in terms of a Ginzburg-Landau-Wilson (GLW) action. 
A next assumption is that the stripe-antiferromagnet orders 
in a {\em continuous} quantum phase transition. This is motivated by
the work of Aeppli {\em et. al.}\cite{aeppli} as discussed in the introduction. 
Leaning heavily on the well understood phenomenology of supersolid
order, together with the work by Zachar {\em et. al.}\cite{zachar}
on the phenomenology
of stripe ordering, I find that the demand for a continuous
transition acts as a strong constraint on the allowed dynamics.
First order behavior is more natural in the present context, and
only under quite specific circumstances second order transitions
can occur. The analysis which follows is not complete. At several
instances a full renormalization group (RNG) analysis is still to be
done, but it is not expected that this will change the picture
radically. Quenched disorder is neglected alltogether. For a
two dimensional order like the stripe phase, quenched disorder
has to dominate eventually\cite{larkin}. However, because the disordering
lengths associated with the static stripe phases tend to be
rather large, it should make sense to analyze first the
clean limit, while disorder physics becomes only of relevance 
very close to the phase transition. This section is organized
as follows: first I will introduce a minimal set of ordering
fields (subsection A). In the absence of
the spin fields, the problem becomes quite
 similar to the problem of supersolids, which will
be discussed next (B). The charge-spin coupling will be discussed,
following the work of Zachar {\em et. al.} (C), and combined with
the supersolid theme in the final subsection (D).

\subsection{The ordering fields.}  

On the level of GLW-theory, the
phase diagram Fig. 1 suggests a rather rich dynamics because
of the involvement of a variety of ordering fields. The order
parameters of relevance are:\\
(i) The spatially uniform d-wave
superconducting orderparameter $\langle e^{i\theta_0} \rangle$,
parametrized in terms of the phase-angle $\theta_0$. The phase
angle $\theta_0$ is conjugate to the uniform charge density 
$N_0$ such that $[ N_0, \theta_0 ] = i$.\\
(ii) The finite wavevector charge density wave order 
$\vec{N}_{2\epsilon}$ associated with the stripe phase charge order.
The total charge density can be written as,
 \begin{equation}
N(x) = N_0 + N_{2\varepsilon, 1} \cos (2\varepsilon x_1) + 
N_{2\varepsilon, 2} \cos (2\varepsilon x_2)
\label{chorpar}
\end{equation}
$2\epsilon$ is the wavevector of the charge order, while the stripe
phase can occur in two orientations ($x_{1,2}$ are the $(1,0)$ and $(0,1)$
directions in the lattice, respectively). The implication is that 
$N_{2\varepsilon}$ is a vector: $\vec{N}_{2\varepsilon} =
(N_{2\epsilon, 1}, N_{2\epsilon, 2})$. As under (i), superfluid
phase angles $\theta_{2\varepsilon, i}$ ($i = 1, 2$)
are conjugated with the charge order, 
corresponding with {\em finite momentum} superconductivity: 
$[ N_{2\varepsilon, i},
\theta_{2\varepsilon, j } ] = i \delta_{ij}$. The interplay of charge
density wave order and superconductivity is the central theme in the
literature dealing with supersolid order.\\
(iii) The novelty is the  incommensurate
antiferromagetic spin order associated with the stripe phase. A
crucial issue is if the spin order is colinear, with the spatial
modulation of the staggered order parameter driven by the magnitude
of the staggered magnetization, or if some spiral modulation is
involved. For the colinear case, the relevant long wavelength theory
is the same as for e.g. a simple two sublattice Heisenberg antiferromagnet
($O(3)$ quantum non-linear sigma model, or the `soft spin' model adapted
here) while the fluctuations of spiral phases are described by more
involved matrix models\cite{spirals}. 
Although direct experimental evidence is not 
available, it is generally believed that the stripe-antiferromagnet
in the cuprates is of the colinear variety, both because this is the
unanimous outcome of theoretical work\cite{capecod}, 
and because of the experience
in the nickelates\cite{tranni}. The staggered spin density is,
 \begin{equation}
\vec{M} (x) =  \vec{M}_{\varepsilon, 1} \cos (\varepsilon x_1) + 
\vec{M}_{\varepsilon, 2} \cos (\varepsilon x_2)
\label{sporpar}
\end{equation}
defining the $O(6)$ rotor field
 $\vec{M}_{\varepsilon} = ( \vec{M}_{\varepsilon, 1},
\vec{M}_{\varepsilon, 2} )$, where $\varepsilon$ refers to the modulation
wavevector, $i = 1, 2$ to the stripe orientation, and
$\vec{M}_{\varepsilon, i} = ( M^x_{\varepsilon, i}, M^y_{\varepsilon, i},
M^z_{\varepsilon, i})$.

\subsection{ Phenomenology of the supersolid.} 

In the absence of spin-order, the
remaining charge sector is  similar to the well studied subject
of supersolid order. Allthough the microscopic physics behind the stripe
phenomenon is clearly quite different from the simple Bose-Hubbard models
discussed in the latter context, there is no obvious reason to expect the
long-wavelength behavior to be different. In the absence of antiferromagnetism,
the progression superconductor-coexistence phase-stripe phase of Fig. 1
translates in the triad superconductor-supersolid-colinear 
charge order known from the study of Bose-Hubbard models\cite{supers,otterlo}. 
Let me recollect some results as of relevance to the present context.    

The starting point is the Bose-Hubbard model,
\begin{eqnarray}
H & = & J \sum_{<ij>} ( a^{\dagger}_i a_j + a^{\dagger}_j a_i ) -
\mu \sum_i n_i + U_0 \sum_i n^2_i \nonumber \\
  & & + U_1 \sum_{<ij>} n_i n_j + U_2 \sum_{<ik>} n_i n_k
\label{boseH}
\end{eqnarray}
where $a^{\dagger}_i$ and $a_i$ are bosonic creation and annihilation
operators obeying $[ a_i, a^{\dagger}_j ] = \delta_{ij}$ ($n_i = 
a^{\dagger}_i a_i$). The parameters $t, \mu$ and $U_0$ are the hopping,
thermodynamic potential and on-site interaction, respectively, while
$U_1$ and $U_2$ are nearest-neighbor and next-nearest-neighbor interactions.
This model has a litteral interpretation in the context of Josephson
junction networks, while in the present context it is no more than a
convenient lattice cut-off model, revealing universal features of the
long wavelength physics.

\begin{figure}[h]
\hspace{0.00 \hsize}
\epsfxsize=0.9\hsize
\epsffile{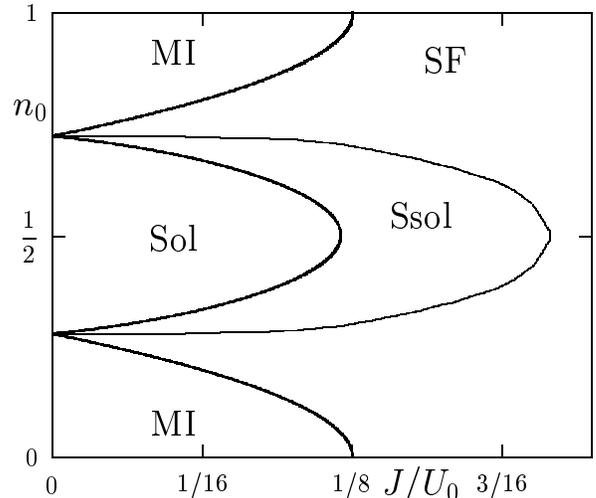}
\caption{ The mean-field phase diagram of the Bose-Hubbard model as function
of $J/U_0$ and average boson density $n_0$, according
to van Otterlo {\em et. al.} [9]. Numerical studies indicate
that the topology of the phase diagram does not change significantly due
to the fluctuations in 2+1D,
for both the checkerboard-[9] and colinear/stripe [8]
charge orders.}  
\label{f3}
\end{figure}

In the absence of the non-local interactions $U_{1,2}$ this model describes
the competition between the condensation of the $q=0$ charge mode 
(Mott-insulator) and the superfluid. For $U_{1,2} \neq 0$ 
charge density wave order is found at particular densities.
If $U_2 \geq U_1$ a particular charge ordering occurs which
is of interest in the present context: a stripe charge order becomes stable
(often called `colinear' in the Bose-Hubbard literature). In Fig. 3
a representative part of the phase diagram is sketched\cite{otterlo},
as function of increasing kinetic energy ($J / U_0$) and average 
particle number ($n_0$) in the grand canonical ensemble, for some
particular choice of non-local interactions. 
At integer fillings ($n_0 = 0, 1, \cdots$) the uniform Mott-insulating (MI)
state is stable for
small kinetic energy, while the stripe state (Sol) acquires stability
at half-integer fillings ($n_0 = 1/2, 3/2, \cdots$). Upon increasing the
kinetic energy, first a phase is entered characterized by a coexistence
of stripe order and superfluidity: the stripe (or colinear) 
{\em supersolid} (Ssol).
Upon a further increase of $J$ the stripe order weakens to disappear at
the phase boundary with the pure superfluid (SF). It is noticed that the
Bose-Hubbard colinear order has much in common with the cuprate stripe order.
For instance, the bond-ordered stripes as found by 
White and Scalapino\cite{whsc} in their
numerical studies of the $t-J$ model (Fig. 4) are quite like the Bose-Hubbard
colinear states assuming that the electrons pair on the elementary plaquet
to form effective bosons\cite{halffilled}. Interestingly, a simple 
explanation is found 
within this framework for the doping independence of the stripe 
wavevector $\varepsilon$ in the doping regime $ 1/8  < x <  0.20$\cite{tran1}.
The stripe phase of the Bose-Hubbard model occurs in the classical limit
($t =0$) only at a half-integer filling, with the associated commensurate
wavevector $\pi / 2a$ ($a$ is the lattice constant). The system would phase separate at non
(half) integer fillings in Mott-insulating and stripe regions. However,
 the supersolid phase can exist
in a homogeneous form away from half integer filling, keeping the wavevector
of the charge order commensurate with the underlying lattice:  
also away from charge commensuration the density wave can stay commensurate
because the excess particle density can be `eaten' by the superfluid order.
Notice that the optimal stability of the stripe phase of Fig. 3 occurs
at half-integer filling; this is quite like the special stability of
the cuprate stripes at the commensurate density $x=1/8$.

A subtle issue is the role played by the finite momentum superconductor,
$\langle e^{i\theta_{2\varepsilon}} \rangle$. In the Bose-Hubbard context
this is playing no role. In fact, by letting the superconductivity
live at $q = 0$ and the charge-order at finite wavevector the either-or
competition is avoided which is a consequence of the number operator being
conjugate to the phase, and this makes possible the existence of the
supersolid. Self-evidently, since translation symmetry is broken by
the charge order, the superconducting order also acquires a spatial
modulation commensurate with the charge order. However, this involves
the {\em amplitude} of the SC order parameter which acquires an admixture
with a finite momentum component. 
However, this component is parasitic and does not play a critical
role. I will assume that this is also the case in the cuprates.

Let us now discuss the nature of the phase transitions of Fig. 3. Obviously,
in the absence of the intervening supersolid phase, the transition between
the stripe phase and the superconductor would be first order. The intervention
of the supersolid, on the other hand, allows in principle for the occurrence
of continuous quantum phase transitions. Although second order transitions
are found on the mean field level,  Frey and Balents\cite{balents}
presented an interesting
analysis showing that the role of critical fluctuations is subtle. For future
use, let me review their arguments. 
The Ginzburg-Landau-Wilson (GLW) action
consistent with the symmetries of  $\vec{N}_{2\varepsilon} =
( N_{2\epsilon, 1}, N_{2\epsilon, 2} )$ (Eq. \ref{chorpar}) is
\begin{eqnarray}
S_N & = & \int d {\bf x} d \tau \{ {1 \over 2} \sum_{i=1}^2 [
( { 1\over {c_N} } \partial_{\tau} N_{2\epsilon,i} )^2 + 
( \nabla N_{2\varepsilon, i} )^2 +
r_N N^2_{2\varepsilon,i} ] \nonumber \\
   &   &   + { {u_N} \over { 4 !} } (\sum_i N^2_{2\epsilon,i} ) )^2 - 
 { {w_N} \over { 4 !} } \sum_i N^4_{2\epsilon,i} ) \} \;,
\label{scharge}
\end{eqnarray}
where $c_N$ is a velocity characterizing the charge order, while
the mass $r_N$ measures the distance from the critical point
associated with the charge ordering. For the quartic anisotropy
parameter  $w_N = 0$ this would correspond (at $k_B T=0$ and 2 space
dimensions) with the GLW action of a classical XY system in
$D=3$. If the anisotropy $w_N > 0$,  stripes oriented along $(1,0)$
or $(0,1)$  are favored in the ordered state.

The superfluid order parameter corresponds with $\langle e^{i \theta_0}
\rangle$, where the superfluid phase $\theta_0$ is governed by the
usual quantum phase dynamics,
\begin{equation}
S_S  =   {1 \over {2 g_S}} \int d {\bf x} d t \left[     
( { 1\over {c_S}} \partial_{\tau} \theta_0 )^2 + ( \nabla \theta_0 )^2\right],
\label{sphase}
\end{equation}
at least deep in the superconducting phase. It is noticed that the 
transition between the supersolid and the stripe phase is actually
governed by a dilute Bose-gas action\cite{fisher} away 
from points of charge commensuration\cite{otterlo,sachdev}.
The physical interpretation is that mobile bosonic defects in
the stripes deconfine and these form initially a dilute gas of bosons. 
Since the stripe phase excitation spectrum is characterized by a 
commensuration gap, the stripe phase orderparameter acts like a spectator at 
this transition.

Of more interest is the transition between the supersolid and the 
superconductor. Because of the massless character of the phase
fluctuations, these can in principle interfere with the critical
fluctuations associated with the charge-ordering transition. The
lowest order allowed coupling between the phase and the charge order
parameter is,
\begin{equation}
S_{NS} =  \int d {\bf x} d\tau i \sigma_{N} ( \partial_{\tau} \theta_0 )
\sum_i N^2_{2\varepsilon, i}
\label{phchco}
\end{equation} 
Frey and Balents\cite{balents} 
show that the critical fluctuations renormalize the
phase velocity $c_S$ (Eq. \ref{sphase}) according to,
\begin{equation}
c^2_{S,R} = { { c^2_S } \over { 1 + const. \xi^{D\alpha/(2-\alpha)} } }
\label{sorenor}
\end{equation}
where $D=3$ (space-time dimensionality) and $\xi$ the correlation length
associated with the stripe ordering. $\alpha$ is the specific heat 
exponent and it is seen from Eq. (\ref{sorenor}) that for $\alpha > 0$
$c^2_{S, R} \rightarrow 0$ at the transition, signalling a runaway
flow, while for $\alpha < 0$ the coupling Eq. (\ref{phchco}) is irrelevant.
It is a classic result of renormalization group theory\cite{brezin}
that the quartic
anisotropy $w_N$ in Eq. (\ref{scharge}) is irrelevant at this transition.
The transition falls therefore in the $D=3$ XY universality class, and
since the specific heat exponent is negative, the coupling to the 
superfluid phase mode is irrelevant as well.

Summarizing, although the direct transition from the superconductor to the
stripe phase is first order, in the presence of a supersolid two continuous
quantum phase-transitions are found: the superconductor-supersolid transition
is a 3D XY transition, and the supersolid-stripe transition is generically
described by the dilute bose gas.

\subsection{ Phenomenology of quantum stripes.}  

As compared to
the previous subsection, the novelty of the cuprate stripe phase is the
prominent role of antiferromagnetism. Neglecting superconductivity, 
the problem remains of the interplay of the finite wavevector charge- and 
spin modes and this has been analyzed on the phenomenological
level by Zachar {\em et. al.}
\cite{zachar}. This work focusses on the finite temperature classical
phase diagram, but it is easily generalized to the 2+1D $k_B T=0$ quantum
dynamics. 

The zero-temperature dynamics of the stripe-antiferromagnetic order
parameter $\vec{M}_{\varepsilon}$ (Eq. \ref{sporpar}) can be represented
by a `soft-spin' GLW action, which is the six-flavor version of 
the charge action, Eq. (\ref{scharge}),
 \begin{eqnarray}
S_M & = & \int d {\bf x} d \tau \{ {1 \over 2} \sum_{i=1}^6 [
( { 1\over {c_M} } \partial_{\tau} M_{\epsilon,i} )^2 + 
( \nabla M_{\varepsilon, i} )^2 +
r_M M^2_{\varepsilon,i} ] \nonumber \\
   &   & + { {u_M} \over { 4 !} } (\sum_i M^2_{\epsilon,i} ) )^2 \nonumber \\
  &   & - 
 { {w_M} \over { 4 !} } ( \vec{M}_{\epsilon,1} \cdot  \vec{M}_{\epsilon,1}
+ \vec{M}_{\epsilon,2} \cdot  \vec{M}_{\epsilon,2} )^2
\} \;.
\label{smagn}
\end{eqnarray}
The quartic anisotropy $w_M$ is choosen such that it leaves the internal
$O(3)$ spin rotation unaffected, breaking the spatial rotation symmetry
to $Z_2$; overall, $O(6)$ is broken by $w_M$ to $O(3)_s \times Z_2$. 
As shown by Brezin {\em et. al.}\cite{brezin}, {\em any} quartic anisotropy
is relevant at the phase transition of a $O(N)$ problem with $N > 4$.
Since $N=6$ for the action Eq. (\ref{smagn}), its phase transition is
governed by $O(3) \times Z_2$ universality. Little attention has been
paid to such symmetry breakings in the statistical physics
literature and the precise nature of its quantum critical regime
is under investigation.

The actions Eq.'s (\ref{scharge},\ref{smagn}) describe the ordering
of the stripe charge- and spin fields independently. Because a fully
developed stripe phase is at the same time charge- and spin ordered,
the mode couplings between these fields should be included. These have
been analyzed by Zachar {\em et. al.}\cite{zachar}. 
Their findings can be directly
applied to the present context of quantum phase transitions. Including
the twofold degeneracy related to the stripe orientation, the lowest
order allowed spin-charge mode couplings are,
\begin{eqnarray}
S_{NM} & = & \int d {\bf x} d \tau \{ {{\lambda_1} \over 2} \sum_{i=1}^2 [
N^*_{2\varepsilon, i} \vec{M}_{\varepsilon,i} \cdot \vec{M}_{\varepsilon, i}
+ h. c. ] \nonumber \\ 
 &  & + {{\lambda_2} \over 2} \sum_{i=1}^2 | N_{2\varepsilon,i} |^2
|\vec{M}_{\varepsilon,i}|^2 \}
\label{chmaco}
\end{eqnarray}
The leading order spin-charge coupling $\lambda_1$ is proportional to
the charge field itself and to the square of the spin-field, because
the former is a scalar and the latter is a vector. This explains directly
why spin orders at the wavevector $\epsilon$ and the charge at $2\epsilon$.
The coupling Eq. (\ref{chmaco}),
together with Eq.'s (\ref{scharge}, \ref{smagn}), defines a 
phenomenological theory for stripe ordering,
\begin{equation}
S_{stripes} = S_N + S_M + S_{NM} \; .
\label{sstripes}
\end{equation}
On the mean-field level, the coupling $\lambda_1$ gives rise to a
rich phase diagram, which is reproduced in Fig. (4).  
Although still to be confirmed by a full RNG analysis, it is expected
that the topology of this phasediagram will not change in three dimensions
if fluctuations are included. This is quite different in two dimensions.
Assuming $2\varepsilon$ to be commensurate with the lattice, and neglecting
the orientational freedom, the charge sector is Ising like
and can therefore order at finite temperature. However, the spin sector carries
a continuous internal symmetry such that magnetic order is forbidden at any
finite temperature according to the Mermin-Wagner theorem. The interpretation
by Zachar {\em et. al.} of the finite temperature phase transitions of LTT cuprates in terms of the phase diagram Fig. 4 was critized by van Duin and
myself\cite{vanduin}. 
We argued that the stripe antiferromagnet is relatively close to the
zero-temperature order-disorder transition, with the effect that the 
2D-3D crossover in the magnetic sector is pushed to low temperatures,
such that mean-field theory looses its validity. 

\begin{figure}[h]
\hspace{0.00 \hsize}
\epsfxsize=0.9\hsize
\epsffile{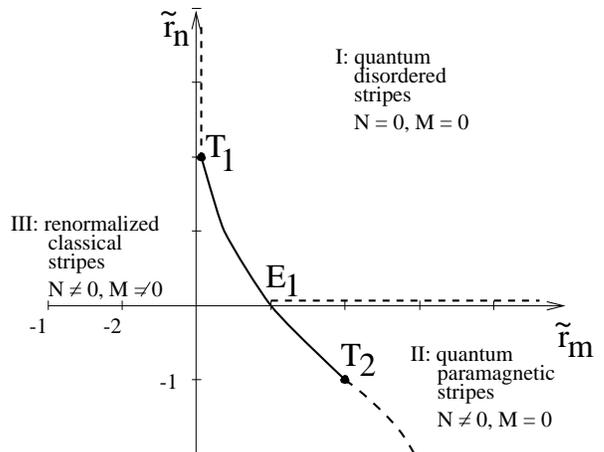}
\caption{ The mean-field phase diagram following from the stripe action
Eq's (\ref{scharge},\ref{smagn},\ref{chmaco}) according to Zachar, Kivelson
and Emery [11], 
here interpreted as a zero-temperature phase diagram. The
axis are the coupling constants of the charge- ($\tilde{r}_n = r_N /
\lambda^2_1$) and spin ($\tilde{r}_m = r_M / \lambda^2_1$)  sectors,
respectively. Dashed lines refer to second order transitions and the heavy 
line corresponds with the spin-charge coupling induced first order transitions.}
\label{f4}
\end{figure}

The quantum ordering dynamics at zero temperature is governed by the (three)
dimensionality of space-time and the topology of the mean-field phase diagram,
Fig. 4, is not expected to be affected by fluctuations in a significant way.
It is therefore expected that a quantum stripe system has the following
phases: (a) Phase I ($r_{M}, r_{N} > \lambda_1^2$): 
the `quantum incompressible stripe phase'. 
Both the spin- and charge sector are quantum disordered. Because the
correlation length in the imaginary time direction is finite in both
sectors, both the charge- and spin excitation spectrum should show  gaps at the
stripe wavevectors. This is the interpretation found in the present
framework for the `dynamical stripes' conjectured to exist in
cuprate superconductors.
(b)  Phase II ( $r_{M}/\lambda^2_1 < 1 $ and/or $r_N/\lambda^2_1 < 2$): the
`renormalized classical stripe phase'. Both spin- and charge are ordered,
and this phase corresponds with the `static' stripe phase. (c)
Phase (III) ( $r_N < 0$ and $r_M/\lambda^2_1$ larger than a critical value):
the `quantum paramagnetic stripe phase'. Although the charge is ordered,
the spin system remains in a quantum disordered state, and is characterized 
by a dynamical mass gap. It is noticed that in principle also a state
can exist which is spin ordered and charge disordered but this involves
necessarily transversal modulations of the spin system (the circular
spiral state of Zachar{\em et. al.}\cite{zachar}).

The phase transitions behave in an interesting way as function of the
various coupling constants. Starting at $r_N >> \lambda^2_1$, there is
a second order transition between the fully disordered state and the
static stripe phase. This transition is driven by the sign change of $r_M$: 
the spin driven transition. The charge mode is massive ($r_N >> 0$) and
is unimportant in the critical regime, as will be further discussed in
subsection D. 

Upon decreasing $r_N$, a regime is entered where the thermodynamics
becomes driven by the spin-charge coupling, Eq. (\ref{chmaco}), and
this causes {\em first order} behavior (heavy line in Fig. 4). Initially,
this first order transition separates the disordered from the fully 
ordered stripe phase, but when $r_N$ changes sign a second order 
charge transition splits off ($E_1$ in Fig. 4). For $r_M > \lambda^2_1$
one finds therefore the sequence: quantum disordered stripes,
quantum paramagnetic stripe phase, and renormalized classical stripe phase.
Initially the spin ordering transition remains first order (due to the
mode coupling) to change to a continuous transition in the purely charge
driven regime. It is noticed that this latter transition is in the 3D
$O(3)$ universality class because the orientational freedom is already
broken at the charge transition.

\subsection{ Stripes and superconductivity: antiferromagnetic supersolids}.

In direct analogy with the coupling between the charge-density mode and the
superfluid phase, Eq. (\ref{phchco}), the coupling between the 
uniform superconductor
and the stripe-antiferromagnet becomes, 
\begin{equation}
S_{MS} =  \int d {\bf x} d\tau i \sigma_{M} ( \partial_{\tau} \theta_0 )
\sum_{i=1}^6 M^2_{\varepsilon, i}
\label{phmaco}
\end{equation}
The crucial observation is that the interplay between finite wavevector
charge order and zero-momentum superconductivity, as discussed in subsection
B, can be `dressed up' with the
stripe antiferromagnetism, without changing the picture drastically.
On the phenomenological level, the magnetic order parameter can be
substituted anywhere for the charge order parameter, with the only
difference that the symmetry is becoming larger. In analogy with the supersolid,
a pure antiferromagnet and a pure superconductor are separated by a
first order boundary. However, a coexistence (antiferromagnetic 
superconductor) phase is thermodynamically allowed and both the
antiferromagnet-coexistence phase and the coexistence phase-superconductor
transitions are of second order. In the context of stripes we meet in
addition the charge-spin mode couplings causing the rich phase diagram,
Fig. 4. Since the charge and spin modes couple in a similar way
to the superconductivity, the supersolid (Fig. 3) and stripe (Fig. 4)
phase diagrams `commute' with each other.

First order boundaries are rather natural in the present context and
I leave it to the reader to enumerate all possible transitions of
this kind. From now on, I insist on the continuous character of the
transition involving the ordering of the stripe antiferromagnet, as
motivated by the observations in Section's I and II. A first condition
is that a coexistence phase should be present; a direct transition from the
singlet cuperconductor to a pure stripe phase is necessarily of first order.
The second condition follows from the stripe phase diagram, Fig. 4: the
charge-spin driven first order transitions should be avoided. By these
simple considerations I find two possible scenario's which allow for
a second order spin ordering transition (Fig. 5). 

\begin{figure}[h]
\hspace{0.00 \hsize}
\epsfxsize=0.9\hsize
\epsffile{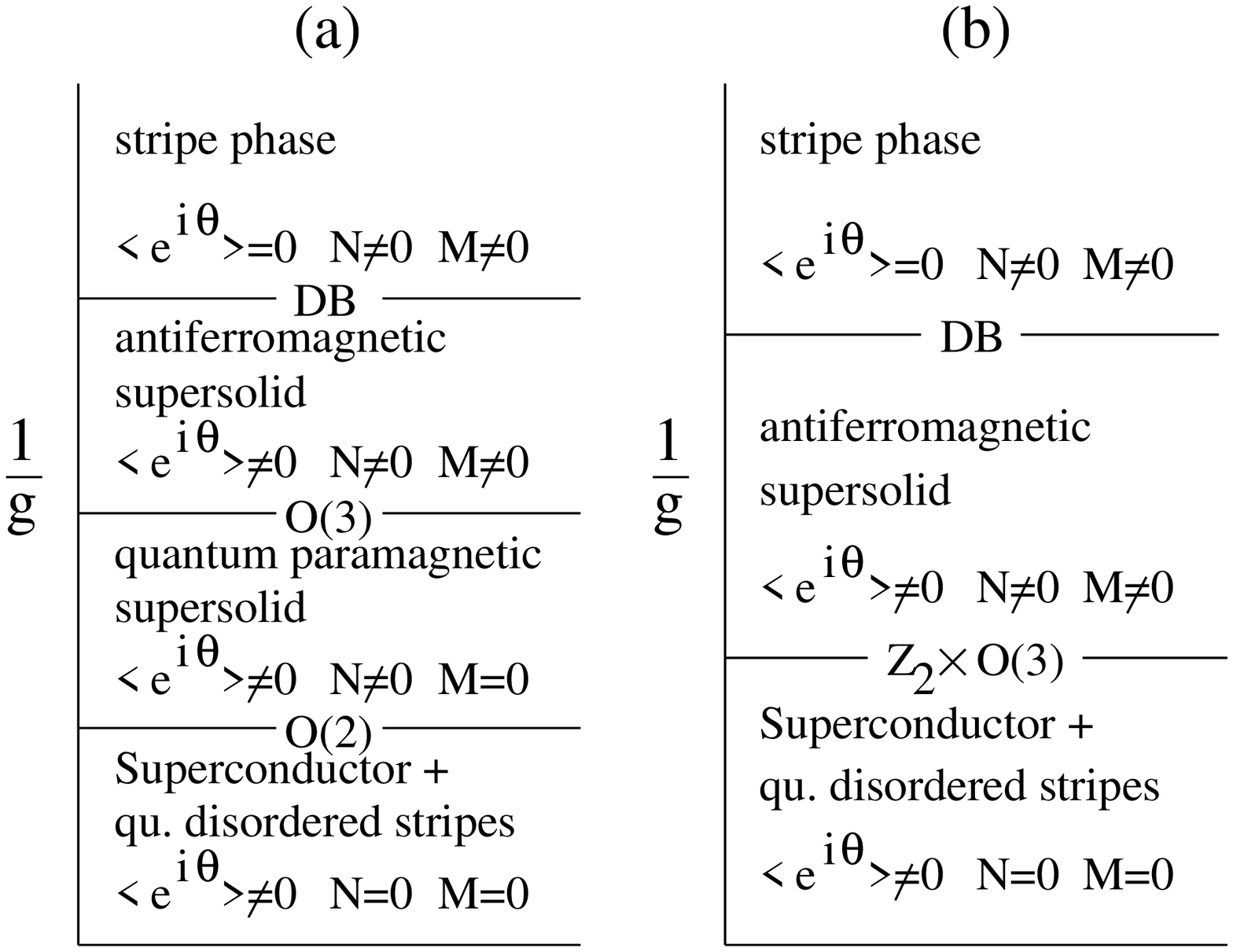}
\caption{ The two possible scenario's, implied by the presence of a 
continuous spin ordering transition. The symmetries governing the
various phase transitions are also indicated (DB is dilute bosons).}
\label{f5}
\end{figure}

{\em Scenario I: independent transitions.} The trivial way to arrive at
a continuous magnetic transition is obviously to let all orderings
occur independently. This is possible if the stripe sector  is in the
`charge driven' (large $\tilde{r}_m$) region of the phase diagram, Fig.
4. A typical sequence of quantum phase transitions as function of
decreasing `$g^{-1}$' could be as indicated in Fig. (5a): the superconductor
acquires a charge order in the $O(2)$ transition of Frey
and Balents\cite{balents}. 
The spin system of this stripe phase is still quantum 
disordered, and orders independently in a standard $O(3)$ transition.
The dilute boson transition where the superconductivity vanishes 
might happen before or after this spin freezing transition; the charge-
and spin sectors are in principle governed by independent coupling
constants and the order in which the transitions happen 
is determined by the microscopy. Assuming that the antiferromagnetic
supersolid exists,  the sequence of the transitions is as
indicated in Fig. 5a. This fingerprint of this scenario is 
a stripe phase which is charge ordered while the spin sector is still
quantum disordered: the quantum paramagnetic supersolid, or in other
words, a superconducting stripe phase with a spin gap. Although such
a state has not been seen in experiments,  it has been (implicitely)
discussed theoretically by Tworzydlo {\em et. al.}\cite{jakub}. 
The scenario Fig. (5a) might appear as less natural for the cuprates. 
It would be
expected that the (quantum) critical fluctuations in the superconducting
state would be dominated by the charge dynamics associated with the
superconductor - paramagnetic stripe phase transition, and not by the
spin fluctations. At the same time, very little is known experimentally
on how the stripe related charge fluctuations behave and this possibility
cannot be excluded on basis of the available data.

{\em Scenario II: The spin-driven stripe ordering.} There is yet another
possibility: the spin-driven regime of the stripe phase diagram, fig. 4.
The phasediagram simplifies in this case (Fig. 5b), and becomes litterally 
like the emperical phase diagram shown in Fig. 1. Different from the
charge driven case (scenario I), the transition is now from the superconductor
directly into the antiferromagnet supersolid. In addition, since the 
transition is dominantly spin driven this possibility appears as more
natural, given the quantum critical spin dynamics observed by Aeppli
{\em et. al.}.

Although one would expect the transition from
a superconductor to an antiferromagnetic supersolid to be of first
order, the transition can be of second order because
the coupling term, Eq. (\ref{chmaco}), can force the
charge fields to follow the spin fields parasitically.
On the ordered side, this implies that the charge orderparameter
grows quadratically slower than the spin order parameter\cite{zachar}.
Defining $r_M \sim (g-g_c) / g_c$ ($g$ is the bare coupling constant and
$g_c$ the critical coupling) and $\beta$ as the order parameter exponent
of the $Z_2 \times O(3)$ transition,  
\begin{eqnarray}
\vec{M} & \sim & \left({ {g_c - g } \over {g_c} } \right)^{\beta} \; ,
\nonumber \\
N   & \sim & \left({ {g_c - g } \over {g_c} } \right)^{2\beta} \; ,
\label{orderpgr}
\end{eqnarray}
a behavior which can easily be checked experimentally by e.g. measuring
the increase of the spin- and charge superlattice peaks as function of
increasing $Nd$ concentration. 

This `slavery' of the charge field to the spin fields 
is also expected to hold in
the quantum disordered regime close enough to the transition. The 
arguments is as follows: in the neighborhood of the spin transition,
where $r_M$ changes sign, the charge sector is still in the disordered
regime, implying a charge-correlation length $\xi_N \sim 1 / \sqrt{ r_N}$
or a charge mass gap $\Delta_N = c_N / \xi_N \sim c_N \sqrt{ r_N}$. 
For lengths $>> \xi_N$ (energies $<< \Delta_N$) these fields can be
integrated out by taking their saddlepoint values. Minimizing $S_{sstripes}$
(Eq. \ref{sstripes}) to the charge fields,
\begin{equation}
N_q = { { \lambda_1 } \over { 4 ( r_N/2  + q^2 ) } } \sum_{i} M^2_{i, q}
\label{slavery}
\end{equation}
including the gradient terms ($q$ is Euclidean momentum). After substitution
of Eq. (\ref{slavery}) in the full action Eq. (\ref{sstripes})
a spin-only action is obtained  with a renormalised quartic term $ u_M / 4! 
\rightarrow u_M / 4! - \lambda_1^2 / ( 2 r_N) $. As long as this quantity
is positive, the critical dynamics is in the spin-only ($Z_2
\times O(3)$) universality class. This implies that the charge-field 
does not carry any dynamics of its own, but follows instead adiabatically
the spin dynamics. This has interesting consequences for the charge dynamics.
Using Eq. (\ref{slavery}) the (dynamical) charge susceptibility becomes
in terms of the euclidean momentum $q$,
\begin{eqnarray}
\chi^N_q & = & \langle N_q N_{-q} \rangle \nonumber \\
     &  \sim & { {\lambda_1^2} \over { (r_N + q^2)^2 } }
\langle ( \sum_{i} M^2_{q,i} ) (\sum_{j} M^2_{-q,j} ) \rangle \;.
\label{chsusc}
\end{eqnarray}
This implies that the stripe-like charge fluctuations will exhibit
a dynamics which is quite similar to the spin dynamics. For instance,
the charge fluctuations will show a quantum gap in the disordered regime
which will be identical to
the spin gap in the magnetic sector. On a more detailed level there will
be differences. On the gaussian level 
$\chi^N_q \sim 1 / (r_N + q^2)^2 (\chi^M_q )^2$ where 
$\chi^M_q = \sum_i \langle M_{i, q} M_{i, -q} \rangle$ (dynamical spin
susceptibility). However, in the 3D case this will no longer be true because 
of the relevancy of the four point vertex.

It is noticed that it remains to be established how the critical fluctuations
associated with this transition interact with the `background' superconductor.
In the charged quasi-2D superfluid, the action Eq. (5) describes the
acoustic plasmon, keeping in mind that the c-axis Josephson plasma
frequency sets  a low energy cut-off.
The arguments by Frey and Balents\cite{balents} for the supersolid 
transition, as discussed
in  subsection B, can now be directly transferred to the 
case of a pure spin transition
(the specific heat exponent $\alpha < 0$ for $O(3)$). 
The subtlety is, however, that the spin ordering is accompanied
by the breaking of spatial rotational symmetry (the two stripe directions),
which changes the universality class of the transition to $Z_2 \times O(3)$
and this has to be studied in further detail.  

Finally, there is a serious problem with this scenario. In the above I asserted
that the zero temperature phase diagram has to do with the spin driven
transition of Zachar {\em et. al.}. At the same time, in the LTT stripe
phases 
the {\em finite} temperature transitions in the stripe ordered region 
of the $T=0$ phase diagram are of the charge driven kind: charge orders
at a higher temperature than the stripe antiferromagnet. At least in the
close neighborhood of the quantum phase transition, where the GLW theory
is valid, such a finite temperature behavior appears as impossible.
In strictly $2+1$ dimensions, any finite temperature will destroy the
spin order, and it is easy to understand that in the realistic case  
(spin anisotropy, $3+1$ D couplings) the spin ordering
temperature can become quite low due to the fluctuations. The problem is,
however, that in the close neighborhood of the quantum transition the charge
sector does not show a tendency to order in the absence of the spins.
In order to find a finite temperature charge ordering transition, it is
necessary to renormalize $r_N$ from a large positive value at $T=0$ to
a negative value at any finite temperature. Since temperature acts in
quantum field theory like a finite size scaling, it is hard to see how this
can happen.

\section{Conclusions}

I have presented here a minimal option for the phenomenological theory
of the zero temperature competition between superconducting- and stripe
order. It is  based on 
current beliefs on the types of order relevant for the cuprates. The
identification of these orders is based on a still highly incomplete
experimental characterization. At the same time, I hope I have convinced
the readership that by elementary considerations a variety of
predictions can be derived.
It is hoped that these issues are taken up by the experimentalists,
who are in the position to prove the above right or wrong.

Let me end this discussion by commenting on some distinct, 
but closely related ideas:
(i) Laughlin argues that the coexistence phase, critical behaviors,
etcetera, are not an intrinsic property of the clean limit but instead
are caused by dirt effects\cite{laughlin}. 
As repeatedly emphasized, first order behavior
is rather natural in the present context. Laughlin argues that 
the `most relevant operator' quenched disorder changes this into a (pseudo)
continuous behavior, while the coexistence phase is a strongly
disordered micro-phase-separated affair of insulating stripes and pure
superconductors. Although this possibility is not excluded, I repeat that
it is not easy to understand how to arrive at the spin quantum criticality 
claimed by Aeppli {\em et. al}. 
 (ii) The quantum liquid crystals
as proposed by Kivelson, Fradkin and Emery\cite{liqcr}. 
There is no conflict between
those ideas and what is presented here. The liquid crystal ideas amount
to the assertion that the charge sector might reveal a substructure
which is more complex than the simple density wave order which has been
considered here. (iii) The `unified' $SO(5)$ ideas of S.C. Zhang\cite{zhang}. 
It is actually the case that the
phenomenology presented here can be completely reformulated in terms
of a  $SO(5)$ action, if appropriate anisotropies are added. For
instance, the antiferromagnetic supersolid can be understood
as a `canted' superspin phase, where the $SO(5)$ vector is canted in a
direction in between the magnetic and superconducting directions
($\pi$ mode condensation). A difference with the  original $SO(5)$
proposal\cite{zhang} is that the antiferromagnetic component is 
now associated with 
the finite wavevector stripe antiferromagnet, instead
of the commensurate magnet of half-filling. Assuming that a mildly broken
$SO(5)$ symmetry is governing the dynamics gives rise to a number of additional
possibilities. For instance, {\em finite momentum} superconductivity appears
as a serious possibility within the $SO(5)$ framework:
 the simplest superconducting stripe phase corresponds with a $SO(5)$
spiral where the superspin rotates from magnetic to superconducting
directions. It follows immediately that the superconductivity lives
at the same wavevectors as the stripe antiferromagnet. Obviously,
the most striking specialty of $SO(5)$ is that the full symmetry
can get restored at isolated point(s) in the zero temperature phase diagram,
such that superconductivity and antiferromagnetism occur on a strictly
equal footing.

Acknowledgements. I acknowledge stimulating discussions with G. A. Aeppli,
V. J. Emery, R. J. Laughlin, S. Sachdev, J. M. Tranquada,
A. van Otterlo, W. van Saarloos, and S.-C. Zhang.

\end{multicols}

\end{document}